\documentclass[12pt]{article}

\usepackage{amssymb}

\font\titlefont=cmbx10 scaled \magstep3

\newcommand{\RR}{{\mathbb{R}}}
\newcommand{\EE}{\mathcal{E}}
\newcommand{\QQ}{\mathcal{Q}}
\newcommand{\ee}{{\sf e}}
\newcommand{\kk}{{\sf k}}
\newcommand{\uu}{{\sf u}}

\def\lprox{\mathrel{\raise .3ex\hbox{$<$\kern-
.75em\lower1ex\hbox{$\sim$}}}}

\def\gprox{\mathrel{\raise .3ex\hbox{$>$\kern-
.75em\lower1ex\hbox{$\sim$}}}}

\begin{document}
\input{epsf}

\begin{center}
{\titlefont ON WORMHOLES WITH\\
\vspace*{0.1in} 
ARBITRARILY SMALL QUANTITIES \\
\vspace*{0.1in}
OF EXOTIC MATTER}\\
\vskip .4in
Christopher J. Fewster\footnote{email: cjf3@york.ac.uk}\\
\vskip .1in
Department of Mathematics,\\
University of York,\\
Heslington,
York YO10 5DD, \\ 
United Kingdom\\
\vskip 0.2in
and
\vskip 0.2in
Thomas A. Roman\footnote{email: roman@ccsu.edu}\\
\vskip .1in
Department of Mathematical Sciences\\ 
Central Connecticut State University\\
New Britain, CT 06050

\end{center}

\centerline{July 4, 2005}

\begin{abstract}
   Recently several models of traversable wormholes have been proposed which 
require only arbitrarily small amounts of negative energy to hold them open 
against self-collapse. If the exotic matter is assumed to be provided by quantum fields, 
then quantum inequalities can be used to place constraints on 
the negative energy densities required. In this paper, we introduce an alternative 
method for obtaining constraints on wormhole geometries, using a recently derived 
quantum inequality bound on the {\it null}-contracted stress-energy averaged 
over a {\it timelike} worldline. The bound allows us to perform a 
simplified analysis of general wormhole models, not just those with small quantities of 
exotic matter. We then use it to study, in particular, the models of 
Visser, Kar, and Dadhich (VKD) and the models of Kuhfittig. The VKD models are 
constrained to be either submicroscopic or to have a large discrepancy between throat size 
and curvature radius. A recent model of Kuhfittig is shown to be 
non-traversable. This is due to the fact that the throat of his wormhole flares outward so slowly 
that light rays and particles, starting from outside the throat, require an infinite lapse of 
affine parameter to reach the throat.

\end{abstract}

\newpage
\baselineskip=14pt
\section{Introduction}
\label{sec:intro}
Recent years have seen much progress in our understanding of the physical laws 
governing negative energy densities associated with quantum fields. It has been 
known for some time that quantum field theory allows 
violations of all the classical pointwise 
energy conditions. In particular, quantum fields violate both the weak energy condition (WEC), which requires
the stress-energy tensor $T_{ab}$ to obey
\begin{equation}
T_{ab}\,u^{a}u^{b}  \geq 0 \,,
\label{eq:WEC}
\end{equation}
for all timelike vectors $u^{a}$, 
and the null energy condition (NEC), which requires
\begin{equation}
T_{ab}\,k^{a}k^{b} \geq 0 \,,
\label{eq:NEC}
\end{equation}
for all null vectors $k^a$ \cite{continuity-comment}. 
Examples are squeezed vacuum states of light \cite{SY} and the Casimir effect \cite{C}, 
both of which can be realized in the laboratory. It is also known that negative energy 
is required for the Hawking evaporation of black holes \cite{H75}, in which an outgoing flux of 
positive energy seen at infinity is paid for by a negative energy flux through the 
horizon \cite{DFU_Can}. 

With this in mind, it is worth considering weaker variants of the WEC
and NEC, based on averages along timelike and null geodesics, 
respectively. Two such conditions are the averaged 
weak energy condition (AWEC):
\begin{equation}
\int_{-\infty}^{\infty} T_{ab}\,u^a u^b \, d\tau \geq 0 \,,
\label{eq:AWEC}
\end{equation}
where $u^a$ is the tangent vector to an inextendible timelike geodesic
parametrized by proper time $\tau$, and the averaged null energy condition (ANEC): 
\begin{equation}
\int_{-\infty}^{\infty} T_{ab}\,k^{a}k^{b} \, d\lambda \geq 0 \,,
\label{eq:ANEC}
\end{equation}
where $k^a$ is the tangent vector to an inextendible null geodesic and $\lambda$ is an affine 
parameter. Violation of the ANEC is known to be a necessary condition for the maintenance 
of traversable wormholes \cite{FSW}.

The fact that quantum field theory allows the existence of states
violating the classical energy conditions raises various concerns. 
If arbitrarily large negative energy densities could persist for
arbitrarily long times,
gross macroscopic effects might occur, including
violations of the second law of thermodynamics or the formation of
exotic spacetime structures. The latter includes ``designer spacetimes''
such as traversable 
wormholes \cite{MT,MTY,VBOOK}, warp drives \cite{A,K98}, and time machines \cite{H92}. 
In two seminal papers, Ford \cite{F78,F91} 
introduced the notion of what have come to be called ``quantum
inequalities'' (QIs) \cite{QEIs}, which 
are restrictions derived from quantum field theory on the magnitude and duration of 
negative energy. More specifically, Ford's original papers were primarily concerned 
with negative energy fluxes. His work was subsequently extended and generalized 
by himself and others to constraints on negative energy densities 
(see Sec.~\ref{sec:QIs} and Refs.~\cite{Few-Rev,TR-MGM10} for recent reviews and more extensive references). 

In this paper we will apply QIs to place constraints on a class of
wormholes introduced Morris and Thorne~\cite{MT} and to some particular
instances recently advanced by Visser, Kar, and Dadhich (VKD)~\cite{VKD} and
Kuhfittig~\cite{KI,KII,KIII}. In so doing, we will make some
improvements to the arguments originally set out in Ref.~\cite{FR96} to
constrain traversable wormhole geometries. The VKD and Kuhfittig models we
study are of interest because they are claimed to use `arbitrarily
small' quantities of exotic matter. In particular, VKD propose a ``volume integral quantifier'' 
which they suggest is a good measure for the amount of exotic matter 
required to maintain a traversable wormhole. Using this measure, they have shown that 
the amount of exotic matter can be made arbitrarily small, even though the ANEC integral 
along radial null geodesics passing through the wormhole is shown to be finite and negative. 
Kuhfittig has also proposed several wormhole models \cite{KI,KII,KIII} 
with similar properties, the last of which he claims to be macroscopic and 
traversable, to require arbitrarily small amounts of exotic matter, and
to be consistent with 
the QI bounds. 

If one assumes that the exotic matter required to maintain these wormholes 
comes from quantum matter fields, then we show, using 
techniques related to those in Ref.~\cite{FR96}, that the geometry of these wormholes is 
severely constrained. However, our analysis differs from that presented in Ref.~\cite{FR96}, 
in that we introduce an alternative 
method for obtaining constraints on wormhole geometries, using a recently derived 
QI bound on the {\it null}-contracted stress-energy averaged 
over a {\it timelike} worldline. The bound allows us to perform a 
simplified analysis of general wormhole models, not just those with small quantities of 
exotic matter. We then use it to study, in particular, the models of 
Visser, Kar, and Dadhich, and the models of Kuhfittig. 

The VKD wormholes 
are constrained by the QI bound to be either submicroscopic in size (e.g., a few orders of 
magnitude above the Planck length), or to have a very small 
ratio of minimum curvature radius to throat size. An examination of Kuhfittig's models 
shows that a confusion between proper 
and coordinate distances in fact renders the model proposed in Ref.~\cite{KIII} {\it non}-traversable. 
In particular, we explicitly show that radially infalling light rays and particles 
reach the throat of this wormhole only after an infinite lapse of affine parameter. 
Lastly, we provide a further justification for our bound using a ``difference inequality'' 
argument.

\section{Quantum inequality constraints on exotic spacetimes}

\label{sec:QIs}
\subsection{Quantum inequalities}

We begin with a short review of quantum inequalities, both to explain
their nature and to set out the
extent of the known results and the classes of states for
which they hold. 

To start, consider $D$-dimensional Minkowski space, and let $\xi(\tau)$ be the 
worldline of an inertial observer with proper time parameter $\tau$ and
velocity $\uu=d\xi/d\tau$. If
$q(\tau)$ is a smooth nonnegative function peaked around $\tau=0$,
with unit characteristic width~\cite{widthnote} and normalized so that 
$\int_{-\infty}^\infty q(\tau)\,d\tau=1$, then the integrals
\begin{equation}
\int_{-\infty}^\infty \langle T_{ab}u^au^b\rangle_\omega(\xi(\tau))
\frac{1}{\tau_0}q(\tau/\tau_0)\,d\tau
\end{equation}
are local averages of the expected renormalized energy density seen over a timescale $\tau_0$
about $\tau=0$ when the field is in state $\omega$. By decreasing
$\tau_0$, we can `zoom in' on the region around $\tau=0$; we may use the
freedom to move the zero
of proper time along $\xi$ to zoom in on different regions of the
worldline. 

Quantum inequalities are constraints on these local averages. 
An example would be a statement of the form: {\em there exists a
dimensionless positive constant
$C$ (depending on $q$ and $D$, but not $\tau_0$ or $\omega$) such that
\begin{equation}
\int_{-\infty}^\infty \langle T_{ab}u^au^b\rangle_\omega(\xi(\tau))
\frac{1}{\tau_0}q(\tau/\tau_0)\,d\tau \ge -\frac{C}{\tau_0^D}
\label{eq:QIform}
\end{equation}
for all physically reasonable states $\omega$ and all sampling times
$\tau_0>0$}. (Here, as elsewhere, we employ units with $\hbar=c=1$.) 
A variety of such bounds have been established in varying
levels of generality and rigor and with varying conditions on the
sampling functions and the class of states involved. The original QIs
(see, e.g., Ref.~\cite{F91,FR95}), were established for the Lorentzian sampling function
$q(\tau)=1/(\pi(\tau^2+1))$, and provided the constant $C=3/(32\pi^2)$
if $D=4$.
A subsequent generalization~\cite{FE,FT} permitted all $q$ of the form 
$q(\tau)=g(\tau)^2$ where $g$ is smooth, real-valued and of compact
support [i.e., vanishing outside a compact set] or of sufficiently rapid
decay at infinity. In this case, for $D=4$,
\begin{equation}
C = \frac{\int_{-\infty}^\infty g''(\tau)^2\,d\tau}{16\pi^2\int_{-\infty}^\infty
g(\tau)^2\,d\tau}\,,
\label{eq:C}
\end{equation}
which has a minimum value of around $5$~\cite{Few-P_Casimir} if $g$ is supported in an
interval of unit length. Note that it is crucial that $g$ be smooth
enough to have a square integrable second derivative~\cite{smooth_note}.

The arguments so far mentioned, i.e., those of Refs.~\cite{FR95,FE,FT}, 
utilize formal manipulations in Fock space and so
are limited---in the first instance---to a class of states arising as
vectors in (or density matrices on) the Fock space built on the Minkowski
vacuum state. These limitations were removed by the first fully rigorous
bound applicable in general dimension $D\ge 2$~\cite{Fewster}. This
bound applies to a general smooth timelike curve $\xi(\tau)$ in a
general globally hyperbolic spacetime and asserts: {\em given any
Hadamard state $\omega_0$ as a `reference state', the bound
\begin{equation}
\int_{-\infty}^\infty \left[\langle T_{ab}u^au^b\rangle_\omega(\xi(\tau))
-\langle T_{ab}u^au^b\rangle_{\omega_0}(\xi(\tau))\right] g(\tau)^2\,d\tau
\ge -\QQ[g]
\label{eq:AGWQI}
\end{equation}
holds for all Hadamard states $\omega$ and smooth, real-valued,
compactly supported $g$, where $g\mapsto\QQ[g]$ is a quadratic form depending on the spacetime, the
trajectory $\xi$, a choice of $D$-bein near $\xi$, and the reference state
$\omega_0$.} It must be emphasized that $\QQ[g]$ is finite for all $g$
in our class, that there is a closed-form expression
for $\QQ[g]$ in terms of the two-point function of $\omega_0$ and, most
importantly, that $\omega$ and $\omega_0$ are {\em arbitrary} Hadamard states:
there is no assumption that $\omega$ can be represented as a vector or
density matrix in the same Hilbert space representation as $\omega_0$
[i.e., $\omega$ and $\omega_0$ may belong to different `folia']. This is because
the argument used in~\cite{Fewster} is formulated within the algebraic
approach to quantum field theory in curved spacetimes, and does not
require the theory to be formulated in Hilbert space. 

The QI, Eq.~(\ref{eq:AGWQI}), is an example of a ``difference QI''; that
is, it bounds the difference of the expectation values of the energy 
density in an arbitrary quantum state and in some reference 
state. However, Eq.(\ref{eq:AGWQI}) is easily converted into an ``absolute QI'' 
\begin{equation}
\int_{-\infty}^\infty \langle T_{ab}u^au^b\rangle_\omega(\xi(\tau))
g(\tau)^2\,d\tau
\ge -\QQ'[g]
\label{eq:AGWQI'}
\end{equation}
where
\begin{equation}
\QQ'[g] =\QQ[g]-\int_{-\infty}^\infty \langle T_{ab}u^au^b\rangle_{\omega_0}(\xi(\tau))
g(\tau)^2\,d\tau
\end{equation}
is also finite for all $g$ in our class, because the integrand on the
right-hand side is smooth and compactly supported. 
Again, it must be emphasized that---contrary to the mistaken view
recently expressed in Ref.~\cite{Kras_v3}---this bound holds for {\em any}
Hadamard state $\omega$, not simply those in the folium of $\omega_0$. 
If we apply the above bounds to the case of an inertial worldline in
four-dimensional Minkowski
space, with $\omega_0$ chosen to be the Minkowski vacuum state then a
bound of the form Eq.~(\ref{eq:QIform}) is obtained if we put
$q(\tau)=g(\tau)^2$ and with $C$ given by Eq.~(\ref{eq:C}).

Let us note a separate strand of work~\cite{FLAN1,Vollick00,FLAN2},
initiated by Flanagan, which treats massless fields in two-dimensional
curved spacetimes for general worldlines and arbitrary Hadamard states. 
There are also various extensions of the QI bounds to 
free Dirac \cite{Vollick00, FewsterVerch, FewsterMistry}, Maxwell
\cite{FR97,Pfenning_em,FewsterPfenning}, Proca \cite{FewsterPfenning} and
Rarita--Schwinger \cite{YuWu} fields. In addition, it has recently been
proved (by extending the argument of Ref.~\cite{FLAN1}) 
that all unitary, positive energy conformal field theories in
two-dimensional Minkowski space obey QI bounds \cite{FewsterHollands},
thus providing the first examples of QIs for interacting quantum field
theories. No general results are known for other interacting quantum
field theories, although Olum and Graham have provided an example in
four-dimensions with two coupled scalar
fields in which a static negative energy density is created. This
suggests that worldline quantum
inequalities might not hold for general interacting quantum field
theories without some further qualification. However, several important
caveats must be entered: first, the Olum--Graham example does not exclude
the possibility that QIs might hold for local averages over suitable spacetime volumes.
We expect that this
is indeed the case--as it is for conformal
fields~\cite{FewsterHollands}--and moreover that this would not
substantially modify the results of our analysis 
in any significant way. Second, the Olum--Graham example
is effectively hard-wired into the Lagrangian, which has been engineered
to produce a domain wall configuration of the required type. It is not
clear to us that a {\em single} choice of Lagrangian (including specific
values for any parameters it contains) could produce {\em arbitrarily} negative static
energy densities. If not, then one might well be able to apply worldline
quantum inequalities on scales shorter than the length scales [implicit in the
Lagrangian] which fix the magnitude of any static negative energy density
configurations. However definite statements on these issues must await more
progress on interacting theories.

\subsection{Constraints on exotic spacetimes}

Quantum inequalities have been used to place constraints on several
different ``designer spacetimes'', such as traversable wormhole and warp drive 
spacetimes \cite{FR96,PFWD,ER,NZK1}. In each case, the basic idea is to obtain
the stress-energy tensor required to support a given spacetime
and then test it for consistency with the QI bounds, leading to
constraints on various parameters arising in the metric. A problem which
must be confronted is that the QI, Eq.~(\ref{eq:AGWQI'}), requires explicit
knowledge of the two-point function of a reference state $\omega_0$ to
compute the right-hand side. Such knowledge is not at hand for general
wormhole models and hinders attempts to use Eq.~(\ref{eq:AGWQI'})
directly. Instead, we will follow Ref.~\cite{FR96} in assuming that the 
{\em flat spacetime} QI bounds should also be 
applicable in curved spacetimes and/or spacetimes with boundaries, in 
the ``short sampling time limit.'' Specifically, we restrict the sampling time to be 
$\tau_0=f \ell_{min}$, where $f\ll 1$ and $\ell_{min}$ is the smallest proper radius of curvature 
or the smallest proper distance to any boundary of the spacetime, and
apply QI bounds for averaging along timelike {\em geodesics}~\cite{Accn_comment}. 

Strictly speaking, this is an {\em assumption}, but it is one for which 
good justification can be provided. Three arguments may be given (see also the
discussion in Ref.~\cite{FR96}): firstly, the equivalence
principle leads us to expect that physics ``in the small'' should be
approximately Minkowskian as far as freely falling observers are
concerned; secondly, it is borne out by specific examples in four
dimensions by taking the short-sampling time limit of various curved
spacetime QIs \cite{FT,PFGQI}; thirdly, one of us (CJF) has recently established 
the validity of this assumption for massless scalar fields in general
two-dimensional spacetimes \cite{Few04}. Further support is provided by
a new argument sketched in Sec.~\ref{sec:DA}. It is also expected that a
more general proof may be given, and work is in progress on this
question. 

The use of flat spacetime QIs in the above fashion suffices to put fairly strong 
constraints on ``designer spacetimes'', such as traversable wormhole and warp drive 
spacetimes \cite{FR96,PFWD,ER}. These analyses were based on QIs
using Lorentzian sampling functions, but the more recent QIs based on 
compactly supported sampling functions 
remove worries that the infinite ``tails'' of a non-compactly 
supported sampling function might invalidate the analysis by picking up
large non-local effects. (This could also be dealt with by making the width of the
sampling function small enough to make 
the sampling function drop off sufficiently fast, at the expense of
weakening the QI bounds.)

\section{Some wormhole geometry}
\label{sec:REVMT}

We study a class of four-dimensional traversable wormholes introduced by Morris and
Thorne \cite{MT} in which two spacetime regions, referred to as the
``upper universe'' and the ``lower universe'' are joined by a throat. 
The wormhole models are static, spherically symmetric, and, for
simplicity, the upper and lower universes are taken to be isometric.
The parameter $\ell$ measures the signed proper radial distance from the
wormhole throat, running from $-\infty$ in the asymptotic region of the
lower universe to $+\infty$ in the asymptotic region of the upper
universe, with $\ell=0$ at the wormhole throat itself. The general form
of the wormhole metric is
\begin{equation}
ds^2=-e^{2\Phi(r(\ell))}dt^2 + {d\ell}^2 
                + {r^2}(\ell)({d\theta}^2+ {\rm sin}^2\theta\,{d\phi}^2)\,,       
                                                \label{eq:MTM2}
\end{equation}
where $2\pi r(\ell)$ is the proper circumference of a circle of fixed
$\ell$ in the equatorial plane $\theta=\pi/2$ (with $t$ constant), 
and $\Phi$ is called the ``red-shift function''. The function
$r(\ell)$ is assumed to be even, twice continuously differentiable, and
to possess a global minimum at the throat, where $r(0)=r_0>0$, and no
other stationary points. It is also assumed that $0<dr/dl\le 1$ for
all $\ell>0$, which ensures that the wormhole
``flares out'' when seen in an embedding diagram, such as the one 
shown in Fig.~\ref{fig:WHprofile}. We also require
that $r(\ell)/|\ell|\to 1$ as $|\ell|\to\infty$, fast enough to ensure
asymptotic flatness. 

The red-shift function $\Phi(r)$ is defined on $[r_0,\infty)$. We
require $\Phi(r(\ell))$ to be twice continuously differentiable and to
satisfy $\Phi(r(\ell))\to 1$ as $|\ell|\to\infty$ fast enough to ensure
asymptotic flatness. Although symmetry between the upper and lower
universes requires
\begin{equation}
\frac{d\Phi(r(\ell))}{d\ell}\biggl|_{\ell=0} = 0 \,,
\end{equation}
we note that $\Phi'(r)$ and $\Phi''(r)$ [i.e., the derivatives with
respect to $r$] may be divergent as $r\to r_0$, 
a point which we will discuss later. 
For the wormhole to be traversable it must have
no horizons, which implies that $g_{tt}=-e^{2\Phi(r)}$ must never
be allowed to vanish, and hence $\Phi(r)$ must be everywhere bounded
from below; it must also be bounded from above by virtue of continuity and
its behavior as $r\to\infty$. 

The restrictions that $\Phi(r(\ell))$ and $r(\ell)$ be twice continuously
differentiable ensure that the stress-energy tensor (obtained from
Einstein's equations) is continuous. It is sometimes useful to weaken
this condition at isolated values of $\ell$ so as to allow the inclusion
of thin shells of matter. 

It is also convenient to introduce a radial coordinate $r$, with range
$[r_0,\infty)$, on the upper universe (or, equally, on the lower
universe) so that the metric now takes the form
\begin{equation}
 ds^2=-e^{2\Phi(r)}dt^2+{{dr^2}\over{(1-b(r)/r)}}
           +r^2({d\theta}^2+ {\rm sin}^2\theta\,{d\phi}^2)\,.           
                                                \label{eq:MTM1}
\end{equation}
Here, $b(r)$, defined on $[r_0,\infty)$, is called the ``shape function'' and
is related to the function $r(\ell)$ by
\begin{equation}
\left(\frac{dr}{d\ell}\right)^2 = 1-\frac{b(r)}{r} \,,
\label{eq:bdef}
\end{equation}
or, equivalently, 
\begin{equation}
\ell=\int_{r_0}^{r(\ell)}{{dr'}\over{(1-b(r')/r')^{1/2}}} \,.
                                                 \label{eq:PD}
\end{equation}
for $\ell>0$. [In the lower universe we would insert an overall minus
sign on the right-hand side.] From Eq.~(\ref{eq:bdef}) we see
that $b(r(\ell))$ must be
continuously differentiable (with a one-sided derivative at $r_0$); 
differentiating Eq.~(\ref{eq:bdef}) and dividing
by $2dr(\ell)/d\ell$ gives
\begin{equation}
\frac{d^2 r(\ell)}{d\ell^2} = \frac{b(r)}{2r^2} - \frac{b'(r)}{2r} \,.
\end{equation}
In particular, as $\ell\to 0$, $b'(r)$ tends to a finite limit
$b_0'=b'(r_0) =1-2r_0 {{d^2 r(\ell)}/{d\ell^2}}|_{\ell=0}$. Note that $b'_0\le 1$, because 
$r(\ell)$ has a minimum at the throat, and therefore\break 
${{d^2 r(\ell)}/{d\ell^2}}|_{\ell=0}\ge 0$.  

Like $r(\ell)$, the shape function $b(r)$ determines the 
outward flaring of the wormhole throat as viewed, for example, in an embedding
diagram; the geometry is completely specified by $\Phi$ together with either $r(\ell)$
or $b(r)$. Since $0\le dr/d\ell \leq 1$ we have $0 \leq 1 -b(r)/r \leq 1$;
since $r(\ell)$ has a unique minimum at $r=r_0$ we see that this is also
the unique solution to the equation $b(r)=r$. Thus $g_{rr}$ diverges at
the throat, but this is clearly only a coordinate singularity as the
metric, Eq.~(\ref{eq:MTM2}), is regular there. We also emphasize that
the proper distance is greater than or equal to
the coordinate distance: $|\ell| \geq r - r_0$.

Substitution of the metric Eq.~(\ref{eq:MTM1}) into the Einstein
equations $G_{ab}=8\pi T_{ab}$ gives 
the stress-energy tensor required to generate the wormhole geometry. In
this section we will use units in which the Planck length is set to
unity. It is also convenient to work in the static orthonormal frame given by the 
basis:
\begin{eqnarray} 
    \ee_{\hat{t}} &=& e^{-\Phi}\,\ee_{t}, \nonumber \\
    \ee_{\hat{r}} &=& (1-b/r)^{1/2}\,\ee_{r}, \nonumber \\
    \ee_{\hat{\theta}} &=& r^{-1}\,\ee_{\theta}, \nonumber \\
    \ee_{\hat{\phi}} &=& (r\,{\rm sin}\theta)^{-1}\,\ee_{\phi}\,,
                                                   \label{eq:SOBASIS}
\end{eqnarray}
where $\ee_t=\partial/\partial t$ etc. 
(These definitions are extended to $r=r_0$ and $\theta=0,\pi$ by continuity.)
This basis represents the proper reference frame of an observer who 
is at rest relative to the wormhole. In this frame the stress tensor 
components are given by
\begin{eqnarray}
T_{\hat t \hat t} &=& \rho 
 ={b' \over {8\pi r^2} } \,, \label{eq:Ttt} \\
 T_{\hat r \hat r} &=& p_r   
 =-{1 \over {8\pi} } \, \biggl[{ b \over {r^3} }-{ {2{\Phi}'} \over r }\,
 \biggl(1-{b \over r} \biggr) \biggr] \,, \label{eq:Trr} \\
 T_{\hat \theta \hat \theta} &=& T_{\hat \phi \hat \phi} = P \nonumber \\ 
&=&\!\!\!\!{1 \over {8\pi} } \,\biggl[{1 \over 2 } \biggl( 
{ b \over {r^3} } - { {b'} \over {r^2} } \biggr) + 
{ { {\Phi}'} \over r } \, 
 \biggl(1- { b \over {2r} } - { b'\over 2 } \biggr) + 
  \biggl(1-{b \over r} \biggr) \, ({\Phi}''+({\Phi}')^2) \biggr] \,.
                                        \label{eq:Tang}      
\end{eqnarray}
The quantities $\rho,\, p_r$, and $P$ are the
mass-energy density, radial pressure, and transverse 
pressure, respectively, as measured by a static observer.  
At the throat of the wormhole, $r=r_0$, these reduce to 
\begin{eqnarray}
{\rho}_0 &=& {{b'_0}\over{8\pi {r_0}^2}}  \,,  \label{eq:rho;r0} \\
p_0 &=& - {1 \over{8\pi {r_0}^2}} \,,      \label{eq:p_r;r0} \\      
P_0 &=& {{1 - b'_0} \over {16\pi r_0}} \, 
\biggl(\Phi'_0 + {1 \over {r_0} } \biggr) \,,  \label{eq:P;r0}
\end{eqnarray}
where $b'_0 = b'(r_0)$ and $\Phi'_0 = \Phi'(r_0)$. 
Note that in taking the limit in Eq.~(\ref{eq:P;r0}), we have implicitly assumed that $\Phi'$ does not 
diverge at the throat. 
 
Given the above definitions, we may now see why the wormhole must
violate the NEC. Let $\kk$ be the null vector $\kk=\ee_{\hat t}+\ee_{\hat r}$. Then,
arguing as in Sec.~11.4 of Ref.~\cite{VBOOK},
\begin{equation}
T_{ab}k^a k^b = \rho+p_r = -\frac{e^{2\Phi}}{8\pi
r}\frac{d}{dr}\left( e^{-2\Phi}\left[1-\frac{b}{r}\right]\right)\,,
\label{eq:Tkkintro}
\end{equation}
which reduces to 
\begin{equation}
T_{ab}k^a k^b = \frac{b_0'-1}{8\pi r_0^2} \,,
\end{equation}
at the throat. Since $b_0'\le 1$, we see that the NEC is violated at the
throat unless $b_0'=1$. If $b_0'=1$, we may argue as follows: the
quantity inside the parentheses in Eq.~(\ref{eq:Tkkintro}) vanishes at
$r=r_0$, but is strictly positive for any $r>r_0$. Therefore, by the
mean value theorem there must a point in $(r_0,r)$ at which the derivative in Eq.~(\ref{eq:Tkkintro}) is strictly
positive, and for which the NEC is therefore violated. Since $r$ was
arbitrary, we have proved that the NEC is violated arbitrarily close to the throat~\cite{MVTnote}.

The curvature tensor components are given by
\begin{eqnarray}
R_{\hat{t}\hat{r}\hat{t}\hat{r}}
&=& \biggl(1- {b \over r} \biggr)\,[{\Phi}''\,+\,{({\Phi}')}^2]\, 
+\, { {\Phi}' \over {2r^2} } \, (b - b'r)  \,,    \label{eq:Rtrtr}  \\
R_{\hat{t}\hat{\theta}\hat{t}\hat{\theta}}
&=& R_{\hat{t}\hat{\phi}\hat{t}\hat{\phi}}
= { {\Phi}'\over r}\,\biggl(1- {b \over r} \biggr)  \,,  
                                            \label{eq:Rttheta} \\ 
R_{\hat{r}\hat{\theta}\hat{r}\hat{\theta}}  
&=& R_{\hat{r}\hat{\phi}\hat{r}\hat{\phi}}  
= {1 \over {2 r^3} } \, (b'r-b) \,,   \label{eq:Rrtheta}  \\
R_{\hat{\theta}\hat{\phi}\hat{\theta}\hat{\phi}}  
&=& {b \over {r^3} } \,.             \label{eq:Rthetaphi}
\end{eqnarray}
All other components of the curvature tensor vanish, except for those 
related to the above by symmetry. At the throat, these 
components reduce to
\begin{eqnarray}
R_{\hat{t}\hat{r}\hat{t}\hat{r}}|_{r_0}
&=& { {\Phi'_0} \over {2r_0} } \, (1 - b'_0)  \,,    \label{eq:Rtrtr|r_0}  \\
R_{\hat{t}\hat{\theta}\hat{t}\hat{\theta}}|_{r_0}
&=& R_{\hat{t}\hat{\phi}\hat{t}\hat{\phi}}|_{r_0}
= 0 \,,                           \label{eq:Rttheta|r_0} \\ 
R_{\hat{r}\hat{\theta}\hat{r}\hat{\theta}}|_{r_0}  
&=& R_{\hat{r}\hat{\phi}\hat{r}\hat{\phi}}|_{r_0}  
= -{1 \over {2 {r_0}^2} } \, (1 - b'_0) \,,   \label{eq:Rrtheta|r_0}  \\
R_{\hat{\theta}\hat{\phi}\hat{\theta}\hat{\phi}}|_{r_0}  
&=& {1 \over {{r_0}^2} } \,.             \label{eq:Rthetaphi|r_0}
\end{eqnarray}  
The limit Eq.~(\ref{eq:Rtrtr|r_0}) depends on the assumption that
$\Phi'$ and $\Phi''$ do not 
diverge at the throat. This will turn out to be an important consideration later 
in our discussion. 

Let the 
magnitude of the maximum curvature component be $R_{max}$. 
Therefore the smallest proper radius of curvature (which is also the 
coordinate radius of curvature in an orthonormal frame) is:
\begin{equation}
r_c \approx {1 \over {\sqrt {R_{max}}} }  \,.  \label{eq:l_c}
\end{equation}
We wish to work in a small spacetime volume around the throat of the 
wormhole such that all dimensions of this volume are much smaller than $r_c$, 
the smallest proper radius of curvature anywhere in the region. Thus, in the
absence of boundaries, spacetime can be considered to be 
approximately Minkowskian in this region, and we can apply a flat spacetime
QI-bound, which we now describe.

\section{A `null-contracted' quantum inequality}\label{sec:nullQI}

The QIs discussed in Sec.~\ref{sec:QIs} (and most QIs
proved to date) are constraints on the
energy density as seen by an observer moving along a (not necessarily
inertial) worldline. However this is not
the only possibility. In Ref.~\cite{FewR}, we proved a QI
which constrains the {\em null}-contracted stress tensor $\langle T_{ab}k^ak^b\rangle_\omega$ of a
free scalar field along
a {\em timelike} worldline, where $k^a$ is a smooth null
vector field. The result takes on a particularly simple form for
massless fields in four-dimensional Minkowski space, with averaging
conducted along an inertial worldline, and for a constant null vector
field $k^a$. Let the worldline be $\xi(\tau)$, parametrized by proper
time $\tau$, and with (constant) four-velocity
$\uu=d{\xi}/d\tau$. Then, as shown in Ref.~\cite{FewR}, we have the
QI
\begin{equation}
\int_{-\infty}^\infty d\tau\,\langle T_{ab}\, k^a k^b\rangle_\omega(\xi(\tau))
g(\tau)^2
\ge -\frac{(k_a u^a)^2}{12\pi^2} \int_{-\infty}^\infty d\tau\,g''(\tau)^2\,,
\end{equation}
for all Hadamard states $\omega$ and any smooth $g$ which is compactly
supported in $\RR$. On the left-hand side, the stress tensor is normal-ordered
with respect to the Minkowski vacuum, which is equivalent to
renormalization according to the Hadamard prescription in this case. One
could also consider non-compactly supported $g$ by taking appropriate
limits using sequences of functions with increasing support. We will not
do this, partly to avoid technical issues concerning the limits, but
mainly because it will not be necessary for our application.

Suppose that we are told that a certain state $\omega$ has $\langle
T_{ab}\, k^a k^b\rangle_\omega(\xi(\tau))\le\EE$ during
$0<\tau<\tau_0$. Applying the QI, we know that
\begin{equation}
\EE\int_{-\infty}^\infty g(\tau)^2\, d\tau \ge \int_{-\infty}^\infty \langle T_{ab}\, k^a k^b\rangle_\omega 
\,g(\tau)^2\,d\tau 
\ge -\frac{(k_a u^a)^2}{12\pi^2}\int_{-\infty}^\infty g''(\tau)^2\,d\tau \,,
\end{equation}
for any smooth $g$ compactly supported in $(0,\tau_0)$. Thus
\begin{equation}
\EE \ge -\frac{(k_a u^a)^2}{12\pi^2}\frac{\int_{-\infty}^\infty g''(\tau)^2\,
d\tau}{\int_{-\infty}^\infty g(\tau)^2\,d\tau}\,,
\end{equation}
(provided $g$ is not identically zero) and we are free to optimize the
right-hand side over the class of allowed $g$'s.
The variational problem may be solved by converting it into an
eigenvalue problem~\cite{Few-P_Casimir,FTii} and leads
to the conclusion that
\begin{equation}
\EE \ge -\frac{C(k_a u^a)^2}{\tau_0^4} \,,
\label{eq:const_bd}
\end{equation}
where $C\approx 4.23$. Note that $\EE$ scales quadratically with $\kk$
by definition, which is why the right-hand side of the bound also has
this dependence. 

Our analysis will be based on the application of
this bound to four-dimensional wormhole spacetimes over short timescales. More precisely,
let $\xi$ be a timelike geodesic with four-velocity $\uu$ and suppose that $\kk$ is
parallel-transported along $\xi$. Then, motivated by the equivalence
principle and the above analysis, we assume that: {\em for any
Hadamard state $\omega$, if $\langle
T_{ab}\, k^a k^b\rangle_\omega(\xi(\tau))\le\EE$ for (at least) a proper duration
$\tau_0$ which is short in comparison with the minimum
length-scale characterizing the geometry, then $\EE$ and $\tau_0$ must be constrained by
Eq.~(\ref{eq:const_bd}).} As mentioned in Sec.~\ref{sec:QIs}
this assumption is supported by various examples and its validity (at
least for the close analogue of energy density, rather than the
null-contracted stress-energy tensor) has 
been proved in the two-dimensional case. 

In our application, the spacetimes in question are static, and the
trajectory $\xi$ will be a static trajectory. If $\omega_0$ is the 
ground state of the quantum field theory on this spacetime, then
$\EE_0=\langle T_{ab}\, k^a k^b\rangle_{\omega_0}(\xi(\tau))$ will be
constant in $\tau$. In examples of static spacetimes where the ground state
stress-energy tensor is known, $|\EE_0|$ is typically two orders of
magnitude smaller than the magnitude of the right-hand side of Eq.~(\ref{eq:const_bd}),
if $\tau_0$ is comparable with the length-scales characterizing the
geometry. Furthermore, one may prove that Casimir energies in locally Minkowskian spacetimes 
are consistent with this requirement~\cite{Few-P_Casimir}. 
So our expectation is that our assumption holds for ground states on
static spacetimes, with considerable room to spare.

This consistency is clearly a {\em necessary} condition for the validity
of our assumption. In Sec.~\ref{sec:DA} we will argue that it is also
a {\em sufficient} condition. Thus we have good reason to believe that
our assumption will produce reliable results.

\section{General analysis}

We begin by examining what conclusions may be drawn on the general
symmetric Morris--Thorne
wormhole model, before passing to particular examples. We initially 
assume only that $b_0'<1$, and discuss the case $b_0'=1$ separately.

Let $\kk=\ee_{\hat{t}} + \ee_{\hat{r}}$. Then $\kk$ is everywhere null,
and parallel-transported along the trajectory
$\xi(\tau) =(e^{-\Phi(r_0)} \,\tau,r_0,\pi/2,0)$,
which is the worldline of a static observer at the throat. Then 
$T_{ab}\, k^a k^b$ takes the constant value
\begin{equation}
\EE=\frac{b_0'-1}{8\pi r_0^2l_p^2}<0 \,,
\label{eq:Tkk}
\end{equation}
along $\xi(\tau)$, where $b_0' = b'(r_0)$ as before and we have reinserted
the Planck length $l_p$ (keeping $\hbar=c=1$) for later convenience. On
the assumption that the stress-energy tensor is generated by a Hadamard
state of a free scalar quantum field, we
therefore have
\begin{equation}
\frac{1-b_0'}{8\pi r_0^2l_p^2} \le \frac{C}{\tau_0^4} \,,
\label{eq:QI-null}
\end{equation}
from Eq.~(\ref{eq:const_bd}), for all $\tau_0$ small compared to local geometric scales. Note that
$k_a u^a =1$ for this trajectory, and also that the
left-hand side is necessarily nonnegative. Let $\ell_{min}$ be
the minimum length scale characterizing the local geometry. Then setting
$\tau_0 = f \ell_{min}$ in Eq.~(\ref{eq:QI-null}) and taking square roots we get
\begin{equation}
\frac{\ell^2_{min}}{r_0} \sqrt{1-b'_0} \le f^{-2}\sqrt{8\pi C} \, l_p\,.
\label{eq:premaster}
\end{equation}
Although we could easily proceed with a general value of $f\ll 1$, we will 
take the more concrete path of fixing $f= 0.01$, which is quite a generous interpretation of
$\tau_0\ll \ell_{min}$. As $\sqrt{8\pi C}\approx 10.3$, and we are only
really interested in order of magnitude estimates, this gives
\begin{equation}
\frac{\ell^2_{min}}{r_0} \sqrt{1-b'_0} \lesssim 10^5 \, l_p\,.
\label{eq:master}
\end{equation}
Clearly, one or both of $\ell^2_{min}/r_0$ or $\sqrt{1-b'_0}$ must be small in
order for this to be satisfied. In fact our assumptions are quite
conservative: as we will see in Sec.~\ref{sec:DA}, violation of Eq.~(\ref{eq:master})
occurs only if the vacuum stress-energy tensor is ten orders of magnitude
larger than its value in typical static spacetimes. It is therefore
likely that the actual constraints on wormholes arising from quantum
field theory are yet stronger than those we describe below. 

Before considering the consequences of this bound, let us note that
our analysis has the following advantage over the one in Ref.~\cite{FR96}.
In the case of wormholes with $\rho \ge 0$ for static observers, it was necessary 
in Ref.~\cite{FR96} to consider the usual QIs applied in the frame 
of a boosted observer passing through the throat in order to get a bound on 
energy density. The greater the boost, however, the shorter the proper
time the observer will spend near the throat, and one should also
consider the transit-time across the region of exotic matter as a
relevant timescale in the analysis. This problem is absent from the present approach,
because we use a {\it null}-contracted 
stress tensor averaged over the timelike worldline of a {\it static} observer 
at the throat. Hence we do not have to worry about the observer 
leaving the region of exotic matter. 

We now begin our analysis of Eq.~(\ref{eq:master}). 
First, assume that $\ell_{min}$ is given by the minimum local curvature
radius $r_c$. An examination of the curvature components 
shows that, ignoring constants of order $1$, the three competing curvature radii at the throat 
are: 
$r_0$, $r_0(1-b'_0)^{-1/2}$ and $(r_0/(|\Phi'_0|[1-b'_0]))^{1/2}$. 
In the last case, we have assumed that $\Phi'$ and $\Phi''$ are non-divergent at the throat. 
Later we will consider what happens if this is not true. 

If the minimum radius of curvature is $\ell_{min}=r_0$, we obtain 
from Eq.~(\ref{eq:master}) that
\begin{equation}
r_0 \lesssim \frac{10^5 \, l_p}{\sqrt{1-b'_0}}  \,.
\label{eq:case_i}
\end{equation}
Macroscopic wormholes in this regime therefore require extreme fine-tuning of
$b'_0$: even to approach a throat radius of $10^{20}$ Planck lengths 
($10^{20}l_p \approx 10^{-15}$m $\approx 1$ fermi $\approx 1$ proton radius) 
one needs $1-b'_0\le 10^{-30}$. 

Next consider the case where $\ell_{min} =
r_0(1-b'_0)^{-1/2}$ is the minimum curvature radius. Equation~(\ref{eq:master}) then becomes
\begin{equation}
r_0 \lesssim 10^5 \, l_p \, \sqrt{1-b'_0} \,.
\label{eq:case_ii}
\end{equation}
Even our wormhole with a small $10^{20}$ Planck length-sized throat clearly requires $b'_0\sim -10^{30}$. 
Note that the curvature radius $\ell_{min} = r_0 / \sqrt{1-b'_0}\lesssim 10^5 \, l_p$ 
[from Eq.~(\ref{eq:case_ii})] 
so one could arguably exclude these wormholes as unphysical, at least for the purposes of traversability.

Let us now consider the case when $(r_0/(|\Phi'_0|[1-b'_0]))^{1/2}$ 
is the smallest local proper radius of curvature, where we continue to 
assume that $\Phi'_0$ is finite at the throat. 
From Eq.~(\ref{eq:master}), we obtain the constraint
\begin{equation}
|\Phi'_0|^{-1} \lprox 10^5 l_p \,\sqrt{1-b'_0} \,,
\end{equation}
which implies a minimum local radius of curvature
\begin{equation}
\ell_{min} \approx \sqrt{\frac{r_0}{|\Phi'_0| (1-b'_0)}}  \lprox  
\frac{\sqrt{10^5l_pr_0}} {{(1-b'_0)}^{1/4}} \,, 
\label{eq:ell_min_3}
\end{equation}
which is roughly $\sqrt{10^5l_p r_0}$ if $b'_0$ is not very close to $1$. 
With this assumption, for a ``human-sized'' wormhole with $r_0 \approx
1\,{\rm m} \approx 10^{35}\,l_p$, 
we have that $\ell_{min} \lprox 10^{20} \, l_p \approx 10^{-15}\,{\rm m}$. 
However, to be {\it traversable} for a human traveller, the wormhole must 
satisfy the radial tidal constraint, 
$|R_{\hat{t}\hat{r}\hat{t}\hat{r}}| \lprox 1/{(10^8 \,\,{\rm m})}^2$, 
(see for example, Eqs. (47a) and (49) of Ref.~\cite{MT}). At the throat this 
reduces to 
\begin{equation}
|R_{\hat{t}\hat{r}\hat{t}\hat{r}}|= \frac{|\Phi'_0| (1-b'_0)}{2r_0} \approx 
\frac{1}{\ell^2_{min}}\lprox \frac{1}{{(10^8 \,\,{\rm m})}^2} \,,
\end{equation} 
in our case, which means that we must have $\ell_{min} \gprox 10^8 \,{\rm m}$. 
Since in the present case, $\ell_{min} < r_0$ by assumption and 
$\ell_{min} \gprox 10^8 \,{\rm m}$ for human traversability, let us set 
\begin{equation}
r_0 = \sigma \, 10^8 \,{\rm m} \,,
\end{equation}
with $\sigma > 1$. If we combine the last expression with Eq.~(\ref{eq:ell_min_3}), 
we get that 
\begin{equation}
\sigma \gprox 10^{38}\, {(1-b'_0)}^{1/2} \,,
\end{equation}
and therefore 
\begin{equation}
r_0 \gprox 10^{46} \, {(1-b'_0)}^{1/2}\,{\rm m} \,.
\end{equation}
Thus we conclude that either $r_0$ is enormous, e.g., if $1-b'_0\sim 10^{-10}$ we have
$r_0 \gprox 10^{41}\, {\rm m} = 10^{25}$ light years, or $b'_0$ is incredibly 
fine-tuned, e.g., $1-b'_0 < 10^{-72}$ for $r_0=10^{10}\, {\rm m}$, i.e., $\sigma=100$.

We now examine the case where $\Phi'(r)$ or $\Phi''(r)$ may diverge at the throat. 
Recall that if the wormhole is symmetric, then we must have 
$d\Phi(r(\ell))/d\ell =0$ at the throat. However, since 
\begin{equation}
\frac{d\Phi(r(\ell))}{d\ell} = \Phi'(r)\sqrt{1-b(r)/r} \,,
\end{equation}
it is possible for $\Phi'(r)$ to diverge at the throat 
without the occurrence of a curvature singularity, provided that 
$\Phi'(r)$ diverges no faster than ${(1-b(r)/r)}^{-1/2}$. Similarly, since 
\begin{equation}
\frac{d^2\Phi(r(\ell))}{d\ell^2} = \biggl(1- \frac{b(r)}{r} \biggr) \,\Phi''(r) 
+\frac{1}{2 r} \, \biggl( \frac{b(r)}{r}-b'(r) \biggr) \, \Phi'(r) \,,
\end{equation}
then at the throat, for $\Phi'(r)$ finite, $\Phi''(r)$ could diverge, provided that 
it diverges no faster than ${(1-b(r)/r)}^{-1}$. Of course both $\Phi'$
and $\Phi''$ could diverge provided their contributions cancel in the
limit. Therefore, we must be careful in interpreting the derivatives of $d\Phi(r)/dr$ 
and $d^2\Phi(r)/dr^2$ at 
$r_0$. One can circumvent these worries by writing the curvature tensor components 
using the metric written in proper radial coordinates, Eq.~(\ref{eq:MTM2}). 
Then one finds, in particular, that \cite{VBOOK-1395}
\begin{equation}
R_{\hat{t}\hat{r}\hat{t}\hat{r}}
= -\frac{d^2 \Phi(r(\ell))}{d\ell^2} \,-\,
{\biggl(\frac{d \Phi(r(\ell))}{d\ell}\biggr)}^2 \,,
\end{equation}
and
\begin{equation}
R_{\hat{t}\hat{\theta}\hat{t}\hat{\theta}}
=
R_{\hat{t}\hat{\phi}\hat{t}\hat{\phi}}=\frac{1}{r(\ell)}\frac{dr(\ell)}{d\ell}\frac{d\Phi(r(\ell))}{d\ell}\,.
\end{equation}
Since $d\Phi(r(\ell))/d\ell=0$ at the throat, $\ell=0$, and
$\Phi(r(\ell))$ is required to have bounded second derivatives with
respect to $\ell$ we have 
\begin{equation}
R_{\hat{t}\hat{r}\hat{t}\hat{r}}\biggl|_{\ell=0}
= -\frac{d^2 \Phi(r(\ell))}{d\ell^2}\biggl|_{\ell=0} \,,
\label{eq:R_{trtr}_ell_r_0}
\end{equation}
and 
\begin{equation}
R_{\hat{t}\hat{\theta}\hat{t}\hat{\theta}}\biggl|_{\ell=0}
= R_{\hat{t}\hat{\phi}\hat{t}\hat{\phi}}\biggl|_{\ell=0}
= 0  \,.
\label{eq:other_Rs_r_0}
\end{equation}  
                                              
If $R_{\hat{t}\hat{r}\hat{t}\hat{r}}$ is not the largest curvature
component then the analysis reduces to one of the cases considered in
Eqs.~(\ref{eq:case_i}) and~(\ref{eq:case_ii}) above. If it is the largest curvature component, then the smallest 
local proper radius of curvature is ${(|d^2 \Phi(r(\ell))/d\ell^2|)}^{-1/2}$. 
From Eq.~(\ref{eq:master}), we then obtain the constraint
\begin{equation}
\ell_{min} \approx {\biggl(\biggl|\frac{d^2 \Phi(r(\ell))}{d\ell^2}\biggr|\biggr)}^{-1/2}
\lesssim \frac{\sqrt{10^5l_p r_0}}{{(1-b_0')}^{1/4}} \,,
\label{eq:ell_min-again}
\end{equation}
which entails the same fine-tuning constraints on $b'_0$ discussed after Eq.~(\ref{eq:ell_min_3})
above.

We conclude this section with some remarks on the case in which
$b'_0=1$. Since the NEC is not violated at the throat, the above
analysis does not apply. However, we have already seen that the NEC is
violated arbitrarily close to the throat, and one could modify the
analysis by considering static trajectories with $r>r_0$ where NEC is
violated. To do so would require more information about the shape and
red-shift functions, and we do not pursue this direction further. It
seems to us that the $b_0'=1$ case is nongeneric, because it corresponds
to a $r(\ell)$ having a minimum at $\ell=0$ with $r''(0)=0$. One would
not expect this nongeneric feature to be stable against 
small fluctuations in the metric (either due to quantum effects, or the
passage of a material body through the wormhole throat). We note that
this criticism could be levelled at some of the Kuhfittig models to be
considered later; a stronger objection is that they fail to be
traversable, as we will see.

\section{The Visser-Kar-Dadhich models}
\label{sec:VKD}
Visser, Kar, and Dadhich (VKD) \cite{VKD} (see also~\cite{NZK2}) have recently suggested 
that a suitable measure of the ``amount of exotic matter required'' for 
wormhole maintenance is given by integrating $\rho+p_r$ (to quantify the
degree of NEC violation) with respect to
the measure $dV=r^2\sin\theta \,dr\,d\theta\,d\phi$ to obtain
\begin{equation}
\int [\rho + p_r]\, dV = 2\int_{r_0}^{\infty}[\rho + p_r]\,4\pi r^2 \,dr \,.
\label{eq:total}
\end{equation}
The factor of $2$ comes from including both wormhole mouths. 
The overall form of Eq.~(\ref{eq:total}) and the integration measure 
are chosen to generalize the mass formula for relativistic stars to 
wormholes \cite{int_measure}. VKD then argue that for a traversable 
wormhole, although the ANEC (line) integral must be 
finite and negative, the volume integral given in Eq.~(\ref{eq:total}) can be made as 
small as one likes. Therefore they conclude that the amount of exotic matter required 
to maintain a traversable wormhole can be made arbitrarily small.

\subsection{\it Spatially Schwarzschild wormhole}
\label{sec:SS} 
VKD introduce two specializations of their model. We will treat each in turn. 
The first is what they call the ``spatially Schwarzschild (SS)'' wormhole. 
They choose 
$b(r) \equiv 2m = r_0$, so that the spatial metric is exactly 
Schwarzschild and the energy density (measured by static observers) $\rho$ 
is zero throughout the spacetime \cite{VBOOK-ZD}. 
In particular, VKD consider a wormhole whose 
metric only differs from Schwarzschild in the region from the throat out 
to some radius $r=a$ (which would have to be reflected in the structure of $\Phi(r)$, 
since $b(r)=const$). They then argue that by considering a sequence of traversable 
wormholes with suitably chosen $a$ and $\Phi(r)$, with $b(r)\equiv 2m=r_0$, 
they can take the limit $a \rightarrow 2m$, and construct traversable wormholes 
with arbitrarily small amounts of exotic matter.

Consider a static observer at the throat of the wormhole. 
For the SS wormhole, $b(r) \equiv 2m = r_0$, and $b'=0$, so
$\rho(r)\equiv 0$. 
Since the energy density is zero in the static frame, to obtain a bound 
using the usual QIs, one would need to boost to the frame 
of a radially moving geodesic observer (see Ref.~\cite{FR96}). The
current approach using the {\it null}-contracted stress energy makes this unnecessary, 
as the radial pressure term is included in $T_{ab} k^a k^b$. From Eq.~(\ref{eq:Tkk}) 
in this case we simply have
\begin{equation}
T_{ab}\, k^a k^b = -\frac{1}{8\pi r_0^2} \,,
\label{eq:TkkSS}
\end{equation}
at the throat. For this wormhole,
the non-zero curvature components are 
\begin{eqnarray}
R_{\hat{t}\hat{r}\hat{t}\hat{r}}
&=& \biggl(1- {r_0 \over r} \biggr)\,[{\Phi}''\,+\,{({\Phi}')}^2]\, 
+\, { {{\Phi}' r_0} \over {2r^2} }   \,,    \label{eq:RtrtrSS}  \\
R_{\hat{t}\hat{\theta}\hat{t}\hat{\theta}}
&=& R_{\hat{t}\hat{\phi}\hat{t}\hat{\phi}}
= { {\Phi}'\over r}\,\biggl(1- {r_0 \over r} \biggr)  \,,  
                                            \label{eq:RtthetaSS} \\ 
R_{\hat{r}\hat{\theta}\hat{r}\hat{\theta}}  
&=& R_{\hat{r}\hat{\phi}\hat{r}\hat{\phi}}  
= -{r_0 \over {2 r^3} } \,,   \label{eq:RrthetaSS}  \\
R_{\hat{\theta}\hat{\phi}\hat{\theta}\hat{\phi}}  
&=& -{r_0 \over {r^3} } \,.             \label{eq:RthetaphiSS}
\end{eqnarray}

Let us first consider the case where $R_{max}=|R_{\hat{t}\hat{r}\hat{t}\hat{r}}|$. 
For the current argument it is simpler to consider this component expressed in terms of proper length. 
If this is the largest curvature component, then the discussion in the last section 
and Eq.~(\ref{eq:R_{trtr}_ell_r_0}) imply that, at the throat, the smallest local 
proper radius of curvature is $(|d^2 \Phi(r(\ell)) / d\ell^2|)^{-1/2}$. 
Then from Eq.~(\ref{eq:ell_min-again}), and the fact that $b'(r)=b'_0=0$ for SS wormholes, 
we have 
\begin{equation}
\ell_{min} \approx {\biggl(\biggl|\frac{d^2 \Phi(r(\ell))}{d\ell^2}\biggr|\biggr)}^{-1/2}  \lprox  
\sqrt{10^5l_pr_0}  \,. 
\label{eq:ell_min_reduced}
\end{equation}
For a ``human-sized'' wormhole with $r_0 \approx
1\,{\rm m} \approx 10^{35}\,l_p$, 
we have that $\ell_{min} \lprox 10^{20} \, l_p \approx 10^{-15}\,{\rm m}$. Even 
a wormhole with $r_0 \approx 1 {\rm A.U.} \approx 10^8 {\rm m}$ would have 
$\ell_{min} \approx 10^{-11} {\rm m}$, which is about one-tenth the radius of a hydrogen atom. 
A somewhat larger wormhole, with $r_0 \approx 1 \,{\rm light \, year} \approx 10^{16} {\rm m}$ 
would still have a local radius of curvature $\ell_{min} \approx 10^{-7} {\rm m}$, which is on 
the order of a wavelength of light. However, recall that as discussed in the last section, 
to be {\it traversable} for a human traveller, the wormhole must 
satisfy the radial tidal constraint, 
$|R_{\hat{t}\hat{r}\hat{t}\hat{r}}| \lprox 1/{(10^8 \,\,{\rm m})}^2$. 
At the throat this 
reduces to $|R_{\hat{t}\hat{r}\hat{t}\hat{r}}|= 
1/{\ell^2_{min}}\lprox 1/{(10^8 \,\,{\rm m})}^2$, 
which means that in our case we must have $\ell_{min} \gprox 10^8 \,{\rm m}$. 
Recall that here $\ell_{min} < r_0$ by assumption, and since 
$\ell_{min} \gprox 10^8 \,{\rm m}$ for human traversability, if we set 
$r_0 = \sigma \, 10^8 \,{\rm m}$ with $\sigma > 1$, we get that $\sigma \gprox 10^{38}$ 
and hence $r_0 \gprox 10^{46} \,{\rm m}$.
Thus we conclude that our bound implies that $r_0$ 
for an SS wormhole must be enormous in order to be traversable for human travellers, 
e.g., $r_0 \gprox 10^{46}\, {\rm m} = 10^{30}$ light years, which is about $10^{20}$ times 
the radius of the visible universe!

If $R_{max} \neq |R_{\hat{t}\hat{r}\hat{t}\hat{r}}|$, then recalling Eq.~(\ref{eq:other_Rs_r_0}),  
we see that the largest curvature component is 
\begin{equation}
R_{max} = { 1 \over {r_0^2} } \,, \label{eq:R_max-SS}
\end{equation}
and so the smallest local proper radius of curvature is 
\begin{equation}
\ell_{min} = r_c \approx r_0 \,.  \label{eq:r_c-SS}
\end{equation}
Applying our QI bound, Eq.~(\ref{eq:QI-null}), with $b'_0=0$, we have
\begin{equation}
r_0 \lesssim 10^5 \, l_p \,,
\end{equation}
where we have, as before, chosen $f \sim 0.01$. 
This is similar to the result obtained for the case discussed at the end of the 
``proximal Schwarzschild'' subsection 
in Ref.~\cite{FR96}.  Therefore, it would seem that 
macroscopic ``spatially Schwarzschild'' wormholes are ruled out or highly constrained 
by the QIs.

\subsection{\it Piecewise $R=0$ wormhole}
\label{sec:PR0}  
As a further specialization, VKD consider a segment of $R=0$ wormhole (zero Ricci scalar) 
truncated and embedded in a Schwarzschild geometry. For $r \in \, (r_0=2m,a)$, they choose
\begin{equation}
{\rm exp}[\Phi(r)]= \epsilon + \lambda \sqrt{1-2m/r} \,, 
\end{equation}
and
\begin{equation} 
{\rm exp}[\Phi(r)]= \sqrt{1-2m/r} \,, 
\end{equation}
for $r \in \, (a,\infty)$, with $b(r)=2m$ everywhere. 
Continuity of the metric coefficients implies that 
\begin{equation}
\lambda = 1 - \frac{\epsilon}{\epsilon_s} \,,
\end{equation}
where $\epsilon_s = \sqrt{1-2m/a}$. There is a thin shell of what VKD call
`quasi-normal' matter at $r=a$. VKD argue that by taking 
suitable limits of $\epsilon,\epsilon_s$, they can make the amount 
of exotic matter required to support the wormhole arbitrarily small. Because this 
is a more detailed example of an SS wormhole, with a specific form given for $\Phi(r)$, 
we can make an even stronger argument for ruling out macroscopic wormholes of this type.

We can write $\Phi(r)$ on $(2m,a)$ as 
\begin{equation}
\Phi(r) = {\rm ln} \biggl[ \epsilon + 
\sqrt{1-2m/r}\biggl(1 - \frac{\epsilon}{\epsilon_s} \biggr) \biggr]\,.
\label{Phi(R=0)}
\end{equation}
Now in this case, although $\Phi(r)$ is well-behaved at the throat, 
$\Phi'(r)$ diverges. This is due to the fact that $r$ is a bad coordinate at 
the throat, and because the divergence of $\Phi'$ 
involves factors of $1-2m/r$. One can see this by examining the derivative of $\Phi$ with respect to 
proper length, $d\Phi/d\ell$, which in fact vanishes 
at the throat.  As a result, in this case the limit of 
$R_{\hat{t}\hat{r}\hat{t}\hat{r}} 
= (1- {r_0/ r})\,[{\Phi}''\,+\,{({\Phi}')}^2]+\, { {\Phi}' / {2r^2} } \, (b - b'r)$, 
as $r \rightarrow r_0=2m$, is {\it not} $\Phi'/{2r_0}$, due to the presence of 
$\sqrt{1- {r_0/ r}}$ terms in the derivatives of $\Phi(r)$. These will result in cancellations between 
terms in $R_{\hat{t}\hat{r}\hat{t}\hat{r}}$. (Similar considerations 
apply to $T_{\hat \theta \hat \theta}$ at the throat.) However, an explicit calculation shows 
that in fact
\begin{equation}
R_{\hat{t}\hat{r}\hat{t}\hat{r}} = R_{\hat{t}\hat{\theta}\hat{t}\hat{\theta}}= 0 \,, 
\,\,\,\,\,{\rm at \,\,\, the \,\,\, throat} \,,
\label{eq:R0101_p-throat_2}
\end{equation} 
in this subcase. 

A similar calculation to that of the general SS wormhole case 
yields similar results. We again find that at the throat, 
$r=r_0=2m$, the smallest local proper radius of curvature is 
\begin{equation}
r_0 \lesssim 10^5 \, l_p \,,
\end{equation}
and so macroscopic piecewise $R=0$ wormholes are ruled out.

Quite apart from the QI arguments we have given, it should also be noted that 
there are some practical difficulties with the VKD models as well. 
The smaller the amount of exotic matter used in these wormholes, 
the closer they are to being vacuum Schwarzschild wormholes. Therefore 
the smaller the amount of exotic matter, the longer it will take an observer 
to traverse the wormhole as measured by clocks in the external universe. 
Perhaps one could counter this by moving the wormhole mouths around. 
In addition, the smaller the amount of exotic matter, the more prone the wormhole is 
to destabilization by even very small amounts of infalling positive matter, 
since this matter will be enormously blueshifted by the time it reaches the throat. 

Barcelo and Visser \cite{BV1,BV} have proposed classical non-minimally coupled scalar fields 
as sources of exotic matter for wormhole maintenance. Since such classical fields 
(if they exist) would not be subject to the QIs, one might hope to circumvent 
the restrictions derived from them. However, the Barcelo-Visser wormholes have some 
problems of their own (see Sec. 5 of Ref.~\cite{TR-MGM10}).

\section{The Kuhfittig models} 
\label{Kuhfittig}
Kuhfittig has written a number of papers which attempt to construct wormholes 
which both satisfy the QIs and which require arbitrarily small amounts of exotic matter. 
We will examine three of these papers, which we will denote as KI \cite{KI}, KII \cite{KII}, 
and KIII \cite{KIII}. 

\subsection{Kuhfittig I}
\label{sec:KI}
In his first model, KI \cite{KI}, Kuhfittig sets
$\Phi(r)\equiv 0$ and defines $b(r)$ in three regions: one near the
throat, $r_0\le r\le r_\epsilon$, in which $b(r)=kr+\epsilon(r)$, 
an intermediary region $r_\epsilon\le
r<r_1$ in which $b(r)=kr$ and an outer region $r>r_1$ in which $b(r)$
becomes constant after a smooth transition near $r_1$. Here, $k<1$ is a
fixed parameter, while $\epsilon(r)$ is a $C^2$ nonnegative function which obeys
\begin{equation}
\epsilon(r_0) = (1-k)r_0 \qquad \epsilon'(r_0)=0 \,,
\end{equation}
so that $b(r_0)=r_0$, and 
\begin{equation}
\epsilon(r_\epsilon)=\epsilon'(r_\epsilon)=\epsilon''(r_\epsilon)=0 \,,
\end{equation}
so the transition at $r_\epsilon$ is $C^2$. Kuhfittig's aim in Ref.~\cite{KI}
was to demonstrate the existence of wormhole models in which the exotic
matter can be confined to an arbitrarily small region: the interval
$(r_0,r_\epsilon)$ in this case. We can use our general analysis to see
what constraints are put on this class of models by QIs.

Since $\Phi\equiv 0$ and $b_0'=k$, examination of Eqs.~(\ref{eq:Rrtheta|r_0}) 
and ~(\ref{eq:Rthetaphi|r_0}) shows that the smallest curvature radius at the throat 
is $r_0$, and therefore we have the constraint
\begin{equation}
r_0 \lprox \frac{10^5 \, l_p}{\sqrt{1-k}}\,.
\label{eq:KI-r_0_bound}
\end{equation}
As already mentioned, this requires significant fine-tuning. 
For a wormhole with a $1\,{\rm m}$ throat, $r_0 = 10^{35}$ Planck lengths, 
Eq.~(\ref{eq:KI-r_0_bound}) requires that $1-k\lprox 10^{-60}$, for example.
Since $b_0'\le 1$ for wormhole models, we see that $k=b_0'$ must be
tuned to a precision of at least one part in $10^{60}$. [We note that
Kuhfittig acknowledges that $k$ might need to be taken close to $1$,
although he does not give estimates.] Fine-tuning of $k$ entails that
various coordinate-independent quantities are also fine-tuned. For
example, the Ricci scalar is
\begin{equation}
R = \frac{2b_0'}{r_0^2} \,,
\end{equation} 
at the throat, and is also tuned to within one part in $10^{60}$ for a
$1\,{\rm m}$ throat which satisfies the bound 
Eq.~(\ref{eq:KI-r_0_bound}). The engineering challenge is yet more severe 
for an Earth-sized throat (one part in $10^{74}$) and barely less daunting 
for a proton-sized throat (one part in $10^{30}$).

We also observe that taking $k$ close to unity means that the wormhole
has extremely slow flaring at the throat. The proper radial distance may
be estimated, for $r<r_1$ by
\begin{equation}
\ell(r) = \int_{r_0}^r \frac{dr'}{\sqrt{1-k-\epsilon(r')/r'}} \ge
\frac{r-r_0}{\sqrt{1-k}} \,,
\end{equation}
and so we see that the coordinate distances $r_\epsilon$ and $r_1$ must
be close to $r_0$ to avoid unfeasibly long traversal times if $\ell(r)$ 
gets too large. Again, this
indicates the necessity for fine-tuning of the model.

\subsection{Kuhfittig II}
\label{sec:KII}
In this paper \cite{KII}, Kuhfittig writes his line element as 
\begin{equation}
ds^2=-e^{2\gamma(r)}dt^2 + e^{2\alpha(r)}dr^2 
                + {r^2}({d\theta}^2+ {\rm sin}^2\theta\,{d\phi}^2)\,.       
                                                \label{eq:KIIle}
\end{equation}
The function $\alpha(r)$ is required to have a vertical asymptote at $r=r_0$: 
${\rm lim}_{r \rightarrow r_0^+} \alpha(r) = + \infty$. By comparing the $g_{rr}$ 
coefficients in Eqs.~(\ref{eq:MTM1}) and~(\ref{eq:KIIle}), we can express $b(r)$ 
in terms of $\alpha(r)$ as follows 
\begin{equation}
b(r) = r(1 - e^{-2\alpha(r)}) \,.
\end{equation}
The choice of the behavior of $\alpha(r)$ is designed to make $b'(r)$ 
close to $1$ near the throat, $r=r_0$, in order to satisfy one of the general 
QI bounds (Eq. (95) of Ref.~\cite{FR96}). This condition on $b'(r)$ implies that 
the embedding diagram will flare out very slowly, a fact which Kuhfittig 
himself recognizes. However, he then claims that this slow flaring need 
not be fatal -- a claim which we will show to be mistaken. 

Kuhfittig then modifies his notation (in a rather confusing way), 
in order to emphasize the behaviour at
the throat, by rewriting the metric as 
\begin{equation}
ds^2=-e^{-2\alpha(r)}dt^2 + e^{2\alpha(r-r_0)}dr^2 
                + {r^2}({d\theta}^2+ {\rm sin}^2\theta\,{d\phi}^2)\,,       
                                                \label{eq:KIIle_2}
\end{equation}
replacing the original $\alpha(r)$ by
$\alpha(r-r_0)$. He also makes the choice 
$\gamma(r) = -\alpha(r)$ for the redshift function, 
and lets this new $\alpha(r)$ diverge at the origin $(r=0)$. 
This has the effect of allowing $\alpha(r-r_0) \rightarrow \infty$ as 
$r \rightarrow r_0$, while keeping $\gamma(r)= -\alpha(r)$ finite in the same limit, 
in order to avoid the appearance of an event horizon. We will use this form of the metric 
for the remainder of this subsection, but will revert to the form Eq.~(\ref{eq:KIIle}) 
in the next subsection, in order to there follow the notation given in Ref.~\cite{KIII}.

The first indication of trouble comes from the evaluation of the proper
radial distance to the throat from any point outside. Since 
$e^{\alpha(r-r_0)} > \alpha(r-r_0)$ for all 
$\alpha(r-r_0)$, we have  
\begin{equation}
\ell(r) = \int_{r_0}^{r} \, e^{\alpha(r'-r_0)} \, dr' > 
\int_{r_0}^{r} \,\alpha(r'-r_0)\, dr'  \,,
\end{equation}
which will diverge if $\alpha(r-r_0)$ diverges fast enough as 
$r \rightarrow r_0$. In particular, $\ell$ is infinite if
$\alpha(r-r_0)>{\rm const}\times(r-r_0)^{-1}$ for all $r$ sufficiently
close to $r_0$. Moreover,
a similar argument (using $e^x\ge x^p/p!$ for each $p=1,2,3,\ldots$ and
any $x>0$) shows that $\ell$ is infinite even for a weak divergence such
as $\alpha(r-r_0)>{\rm const}\times (r-r_0)^{-1/p}$ for some $p>0$. More can be said if
$\alpha(r-r_0)$ is monotonically decreasing in some interval
$(r_0,r_\epsilon)$. In this case, $\ell(r)$ is finite and tends to $0$
as $r\to r_0$ only if 
\begin{equation}
0\le (r-r_0) e^{\alpha(r-r_0)} \le \int_{r_0}^{r} \, e^{\alpha(r'-r_0)} \,
dr' \to 0 \,,
\end{equation}
in this limit, where the central inequality holds because
$e^{\alpha(r'-r_0)}\ge e^{\alpha(r-r_0)}$ for $r'\in(r_0,r)$. It
follows that $\alpha(r-r_0)$ is less than
$-\log(r-r_0)$ for all $r$ sufficiently close to $r_0$ [indeed,
$-\log(r-r_0)-\alpha(r-r_0)\to\infty$ as $r\to r_0$]. Certainly the
particular choices employed in Sec. III.A of KII, namely
\begin{equation}
\alpha(r)=\kappa/r\,,\quad r_0=0\,,\quad\kappa=0.00025~\hbox{light years}\,,
\end{equation} 
lead to an infinite proper distance from any point $r_1 > 0$ 
to the throat $r_0=0$. We note that Kuhfittig claims a finite traversal time
for this model at the end
of his Sec. III.A; this appears to be based on a confusion between
coordinate and proper distance.

\subsection{Kuhfittig III}
\label{sec:KIII}
In paper \cite{KIII}, Kuhfittig suggests that 
good choices for $\gamma(r),\,\alpha(r)$ in Eq.~(\ref{eq:KIIle}) are:
\begin{equation}
\alpha(r) = \frac{k}{(r-r_0)^n} \,,\,\,\,\,\,\, n \geq 1 \,,
\label{eq:alpha-gen}
\end{equation}
where, in this subsection, $k$ is a (positive) constant with the same units as $r^n$. 
The choice of $n \geq 1$ 
is made in order to obtain $b'(r) \sim 1$ near $r=r_0$. The function $\gamma(r)$ 
is chosen to be:
\begin{equation}
\gamma(r) = -\frac{L}{(r-r_2)^n} \,,\,\,\,\,\,\, n \geq 1 \,,
\label{eq:gamma-gen}
\end{equation}
where $L$ is another positive constant with the same units as $r^n$, and $0<r_2<r_0$. 
The condition on $r_2$ is made to avoid an event horizon at the throat $r_0$. 

Once again, we see that there is a problem when one evaluates the proper distance 
to the throat from any point outside. As in KII, since $e^{\alpha(r)} > \alpha(r)$ for all 
$\alpha(r)$, we have  
\begin{equation}
\ell = \int_{r_0}^{r_1} \, e^{\alpha(r)} \, dr > \int_{r_0}^{r_1} \,\alpha(r)  \, dr = 
\int_{r_0}^{r_1} \, \frac{k}{(r-r_0)^n}\, dr \,,
\end{equation}
which diverges if $n \geq 1$. Hence the proper distance from any point $r_1 > r_0$ 
to the throat $r_0$ is infinite. Now it is known that there are black hole 
spacetimes where the proper distance to the horizon is infinite, but it nevertheless takes 
only a {\it finite} proper {\it time} to fall into them. An example is the extreme $Q=M$ 
Reissner-Nordstr\"{o}m black hole \cite{Carter,TR88}. In that case, however, 
there is a horizon in the spacetime, so an infalling observer who crosses 
the horizon cannot get back out. By contrast, for a traversable wormhole, it should 
be possible for two static observers on opposite sides of the wormhole to stretch 
a measuring tape between them and measure their separation in proper distance. 
If the proper distance from any observer's location to the throat is infinite, this 
will of course not be possible. 

Let us pursue this reasoning further and calculate the proper time for a 
radially infalling observer to reach the throat. 
For a radial timelike geodesic in this wormhole metric we have that
\begin{equation}
\frac{d^2 r}{d\tau^2} + e^{-2[\alpha(r)-\gamma(r)]} \, \gamma'(r) \, 
{\biggl(\frac{dt}{d\tau} \biggr)}^2 + \alpha'(r)\, 
{\biggl(\frac{dr}{d\tau} \biggr)}^2 = 0 \,,
\label{eq:r-geod}
\end{equation}
where $\tau$ is the observer's proper time.
From the four-velocity $u^a$, we have $u^a u_a=-1$, and thus 
\begin{equation}
-e^{2\gamma(r)} \, 
{\biggl(\frac{dt}{d\tau} \biggr)}^2 + e^{2\alpha(r)} \, 
{\biggl(\frac{dr}{d\tau} \biggr)}^2 = -1 \,.
\end{equation}
If we solve this equation for ${(dt/d\tau)}^2$, and substitute 
back into Eq.~(\ref{eq:r-geod}), we obtain 
\begin{equation}
\frac{d^2 r}{d\tau^2} + \gamma'(r) \, e^{-2\alpha(r)} \, +\,
[\gamma'(r)+ \alpha'(r)]\, 
{\biggl(\frac{dr}{d\tau} \biggr)}^2 = 0 \,.
\label{eq:r-geod_2}
\end{equation} 
This can be solved exactly for any $\alpha(r), \gamma(r)$. 
It has a first integral
\begin{equation}
\frac{dr}{d\tau} = - e^{-[\gamma(r)+\alpha(r)]} \sqrt{K-e^{2\gamma(r)}} \,,
\label{eq:dr/dtau}  
\end{equation}
where $\gamma(r), \alpha(r)$ are evaluated at $r(\tau)$ on the righthand side, 
and $K$ is a constant which fixes the initial radial velocity; the
overall minus
sign on the right-hand side corresponds to initially 
in-going geodesics. (To check this, simply
differentiate both sides with respect to $\tau$ and substitute into Eq.~(\ref{eq:r-geod_2}), 
using Eq.~(\ref{eq:dr/dtau}) again to simplify.) Suppose the initial radius is $r_1$ 
at time $\tau=0$. Then
\begin{equation}
\tau = \int_{r}^{r_1} \,dr' \, \frac{e^{[\gamma(r')+\alpha(r')]}}{\sqrt{K-e^{2\gamma(r')}}} \,,
\end{equation}
is the proper time of (first) arrival at radius r. In the KIII model, $\gamma(r)$ is 
well-behaved at $r=r_0$, but $\alpha(r)$ has a nonintegrable singularity
there. As $\alpha(r)>0$ we deduce that $\exp(\alpha(r))$ also has a
nonintegrable singularity at the throat. Thus 
$\tau \rightarrow \infty$ as $r \rightarrow r_0$, and so the proper time for a radially 
infalling observer to reach the throat is infinite. 

Lastly, let us consider radially infalling light rays. Null geodesics obey 
\begin{equation}
g_{ab} k^a k^b = -e^{2\gamma(r)}\,
{\biggl(\frac{dt}{d\lambda} \biggr)}^2 + e^{2\alpha(r)}\,
{\biggl(\frac{dr}{d\lambda} \biggr)}^2 = 0 \,,
\label{eq:nullgeod}
\end{equation}
where $\lambda$ is an affine parameter. Thus 
\begin{equation}
{\biggl(\frac{dt}{dr} \biggr)}^2 = e^{2 \beta(r)} \,,
\end{equation} 
where $\beta(r)=\alpha(r)-\gamma(r)$. Define $B(r)$ such that 
$B'(r)=e^{\beta(r)}$, i.e., 
\begin{equation}
B(r) = \int_{r_i}^{r} e^{\beta(r')}dr' \,,
\end{equation} 
for some $r_i > r_0$.
Now define radial null coordinates by 
\begin{eqnarray}
u&=&t-B(r) \\
v&=&t+B(r) \,.
\end{eqnarray}
Then we can write
\begin{eqnarray}
du\,dv&=&(dt- e^{\beta(r)}\,dr)(dt+ e^{\beta(r)}\,dr) \\
&=&dt^2-e^{2 \beta(r)} dr^2 \\
&=&-e^{-2 \gamma(r)} ds^2 \,,
\end{eqnarray}
where $ds^2$ is the wormhole metric in the $t,r$ plane, which may then be written as 
\begin{equation}
ds^2 = -e^{2 \gamma(r)}du\,dv \,, \,\,\,\,\,\,\,\,\,\,\,\,\,{\rm with}\,\,
r=B^{-1} \biggl(\frac{v-u}{2} \biggr) \,.
\end{equation}
For Kuhfittig's metric, $\int_{r}^{r_i} e^{\beta(r')}dr' \rightarrow \infty$ as 
$r \rightarrow r_0$, so we have $B(r) \rightarrow -\infty$, as $r \rightarrow r_0$. 
To determine an affine parameter, we use the fact that 
${(\partial /\partial t)}^a$ is a Killing vector, 
so $E=-g_{ab} k^a {(\partial /\partial t)}^b$ is a constant 
along null geodesics \cite{Wald}. So we have that $E={\rm exp}[2 \gamma(r)] dt/d\lambda=
(1/2){\rm exp}[2 \gamma(r)] du/d\lambda$, since $u=2t-v$, and $v=$const on the 
ingoing null rays. Therefore a suitable affine parameter is
\begin{equation}
\lambda(u)= \frac{1}{2E} \int_{u_i}^{u} \,{\rm exp}\,\biggl[
2 \gamma\biggl(B^{-1} \biggl(\frac{v-u}{2} \biggr) \biggr) \biggr] \,du \,.
\end{equation} 
As the throat is approached, $u \rightarrow \infty$, $(v-u)/2 \rightarrow -\infty$, 
so $B^{-1}([v-u]/2) \rightarrow r_0$. If $\lim_{r \rightarrow r_0} \, \exp(2\gamma(r))$ 
is nonzero, as in Kuhfittig's example from Eqs.~(\ref{eq:alpha-gen}) 
 and~(\ref{eq:gamma-gen}), then $\lambda(u) \rightarrow \infty$ as 
$u \rightarrow \infty$, so ingoing radial null geodesics do not arrive at the throat 
at finite affine parameter. Hence even light rays cannot traverse the
wormhole. 

\begin{figure}
\begin{center}
\leavevmode\epsfysize=5cm\epsffile{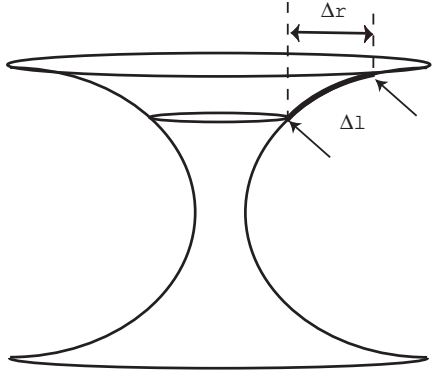}
\end{center}
\caption{Proper versus coordinate distance in a wormhole.}
\label{fig:WHprofile}
\end{figure}

\begin{figure}
\begin{center}
\leavevmode\epsfysize=7cm\epsffile{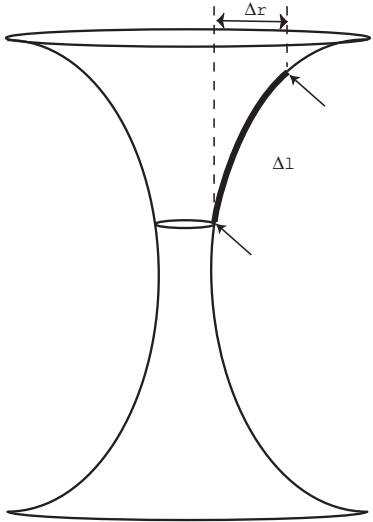}
\end{center}
\caption{Proper versus coordinate distance in a slowly flaring wormhole.}
\label{fig:WHprofile_flared}
\end{figure}

A common problem in all the Kuhfittig models is the confusion of 
coordinate distances and proper distances, as for example, 
in his estimates of traversability times. 
For a slowly flaring wormhole, a small difference in coordinate length 
can correspond to an enormous difference in proper length 
(see Figures~\ref{fig:WHprofile} and ~\ref{fig:WHprofile_flared}). 
As a result, although such a wormhole might have its exotic matter 
concentrated in a small coordinate thickness in radius, the proper volume 
of the region of exotic matter could in fact be very large. 

\section{A further justification for our wormhole bounds}
\label{sec:DA}

At the end of Sec.~\ref{sec:nullQI}, we argued that the Minkowski space
bound Eq.~(\ref{eq:const_bd}) could be adapted to curved spacetimes
under certain hypotheses. In the static case, we noted that a necessary
condition for the validity of this approach is that the bound is
satisfied for the expected stress-energy tensor of the static ground
state (assuming this is Hadamard), with $\tau_0$ of the order of the
minimum length scale characterizing the geometry. Here, we show that this is also a
sufficient condition; this may also be regarded as providing a 
second argument in favor of a bound of 
the form Eq.~(\ref{eq:master}). 
Instead of applying the Minkowski bound on sufficiently small
scales, we may consider what sort of QI could be derived directly in the
wormhole spacetime. In fact our analysis applies to any static globally
hyperbolic spacetime, provided the quantum field theory does not have
bad infra-red behaviour. Suppose a scalar quantum field of mass $m\ge 0$ 
admits a Hadamard static ground state $\omega_0$. Applying Theorem~III.1 in
Ref.~\cite{FewR} to averages along a static trajectory $\xi(\tau)$,
and using arguments similar to those in Sec.~5 of Ref.~\cite{Fewster} we
obtain a bound 
\begin{equation}
\int \left(\langle T_{ab}\, k^a k^b\rangle_\omega(\xi(\tau))-\langle T_{ab}\, k^a
k^b\rangle_{\omega_0}(\xi(\tau))
\right) g(t)^2\,dt \ge -\frac{1}{\pi}\int_0^\infty Q(y)|\widehat{g}(y)|^2\,dy \,,
\label{eq:Qbound}
\end{equation}
for any Hadamard state $\omega$ of the scalar field, where the hat
denotes Fourier transform~\cite{fnote}. As before $g$ is
smooth, real-valued and compactly supported in $\RR$. {\it The advantage
here is that we now no longer need to restrict the support 
of $g$ to be small in relation to curvature scales.}
By arguments parallel to those in Ref.~\cite{Fewster}, the function $Q$ is
non-negative, continuous from the left, increasing, and growing no
faster than polynomially at infinity. In fact, if the two-point function
of the ground state is given by a sum (or integral) of mode functions
\begin{equation}
\langle\varphi(t,x)\varphi(t,x')\rangle_{\omega_0} =
\sum_\lambda e^{-i\omega_\lambda (t-t')}U_\lambda(x) \overline{U_\lambda(x')} \,,
\end{equation}
(note that the $\omega_\lambda$ will be nonnegative if $\omega_0$ is a
ground state) then
\begin{equation}
Q(y) =\sum_{\lambda~{\rm s.t.}~\tilde{\omega}_\lambda<y} |c_\lambda|^2 \,,
\end{equation}
where $\tilde{\omega}_\lambda=e^{-\Phi(r_0)}\omega_\lambda$ and
\begin{equation}
c_\lambda =\left. k^a\nabla_a e^{-i\omega_\lambda t}U_\lambda(x)\right|_{(t,x)=\xi(0)}\,.
\end{equation}
(See, e.g., the introduction to Ref.~\cite{Fewster}. This result could also
be obtained using the less rigorous methods of Ref.~\cite{FT}.)

Suppose that the static spacetime geometry is supported by a particular Hadamard state $\omega$.
Then $\langle T_{ab}\, k^a
k^b\rangle_\omega$ must be constant on the static trajectory $\xi$, as must
the vacuum energy $\langle T_{ab}\, k^a k^b\rangle_{\omega_0}$ because
$\omega_0$ was assumed to be static. These terms may therefore be
taken outside the integral in Eq.~(\ref{eq:Qbound}), and the QI then implies
\begin{equation}
\langle T_{ab}\, k^a k^b\rangle_\omega\ge \langle T_{ab}\, k^a
k^b\rangle_{\omega_0}
 -\frac{\int_0^\infty Q(y)|\widehat{g}(y)|^2\,dy}{\pi\int_{-\infty}^\infty g(t)^2\,dt} \,,
\end{equation}
where the expectation values are evaluated, for example, at $\xi(0)$. 

It is convenient to rewrite the denominator in the last expression in
terms of the Fourier transform, using Parseval's theorem. Moreover,
$|\widehat{g}(y)|^2$ is even in $y$ because $g$ is real-valued, which permits us to write
\begin{equation}
\langle T_{ab}\, k^a k^b\rangle_\omega\ge \langle T_{ab}\, k^a
k^b\rangle_{\omega_0}
 -\frac{\int_0^\infty Q(y)|\widehat{g}(y)|^2\,dy}{\int_{0}^\infty |\widehat{g}(y)|^2\,dy} \,.
\end{equation}
We may now try to maximize the expression on the right-hand side over
the class of $g$ at our disposal.
Fix any compactly supported real-valued $g$ for which the denominator above is unity, and then replace
$g$ by $\lambda^{-1/2}g(\tau/\lambda)$, and therefore $\widehat{g}(y)$ by
$\lambda^{1/2}\widehat{g}(\lambda y)$. As the sampling time is increased
by increasing $\lambda$, $\lambda^{1/2}\widehat{g}(\lambda y)$ becomes more sharply peaked near
$y=0$ ($\widehat{g}(y)$ decays rapidly at infinity because $g$ is
smooth and compactly supported). Taking the limit
$\lambda\to\infty$ (cf. an argument in the proof of
Theorem 4.7 in Ref.~\cite{FVpassive}) we obtain
\begin{equation}
\langle T_{ab}\, k^a k^b\rangle_\omega\ge \langle T_{ab}\, k^a
k^b\rangle_{\omega_0} -Q(0+)\,,
\end{equation}
where $Q(0+)=\lim_{y\to 0^+}Q(y)$. Now $Q(0+)$ will vanish unless the
quantum field theory has bad infra-red behaviour, e.g., a
square-integrable zero mode. Excluding such pathological cases, we have
\begin{equation}
\langle T_{ab}\, k^a k^b\rangle_\omega\ge \langle T_{ab}\, k^a
k^b\rangle_{\omega_0}\,.
\label{eq:DAbd}
\end{equation}
Thus the ground state yields the lowest constant value of
$\langle T_{ab}\, k^a k^b\rangle$ possible (amongst Hadamard states) along
a static trajectory. Accordingly, if the ground state obeys a bound of
the form Eq.~(\ref{eq:const_bd}) with $\tau_0$ of the order of the
minimum length scale characterizing the geometry, then the bound will
hold for all Hadamard states capable of supporting a static geometry. 

Let us develop this line of reasoning a bit further, returning to the
particular class of Morris--Thorne wormholes. In our $\hbar=c=1$
units, the right-hand side of Eq.~(\ref{eq:DAbd}) has the dimensions of $({\rm length})^{-4}$,
and can be written as $K_0 (k_a u^a)^2/\ell_{min}^4$, where $\ell_{min}$ 
is the minimal length scale associated with the geometry.
We know that the wormhole models violate the NEC at [or arbitrarily close to] the
throat, so if $K_0>0$ the wormhole would be inconsistent with QIs. So let us assume
that $K_0<0$. Inserting the required stress-energy tensor, we now have a
bound
\begin{equation}
\frac{b_0'-1}{8\pi r_0^2l_p^2} \ge -\frac{|K_0|}{\ell_{\rm min}^4}
\end{equation}
or
\begin{equation}
\frac{\ell^2_{min}}{r_0} \sqrt{1-b'_0} \le \sqrt{8\pi |K_0|} \, l_p\,.
\end{equation}
which should be compared with Eq.~(\ref{eq:master}). Since the
dimensionless constant $|K_0|$ is
typically of the order of $10^{-2}$, this new bound is stronger than
that of Eq.~(\ref{eq:master}) by five orders of magnitude. This supports
our earlier argument and also suggests that the bounds given above
are extremely conservative: the Casimir
energy would have to be ten orders of magnitude higher than our typical
experience (i.e., $|K_0|\sim 10^8$) in order for Eq.~(\ref{eq:master})
to be violated.

\section{Summary}
\label{sec:summary}

We have analyzed the recent wormhole models of Visser, Kar, and Dadhich, 
and of Kuhfittig. In these models only arbitrarily small amounts of 
exotic matter are required to hold the wormholes open. If the exotic matter is composed 
of quantum fields, then they are subject to the constraints derived from 
the quantum inequality bounds on negative energy. In particular, our analysis 
employs a recently derived quantum inequality bound 
on the {\it null}-contracted stress-energy averaged over a {\it timelike} worldline. 
The bound allows us to perform a simplified analysis of general wormhole models, 
not just those with small quantities of exotic matter.

We showed that our bound implies that {\it macroscopic} wormholes of the Visser-Kar-Dadhich type are 
ruled out or severely constrained. For the Kuhfittig models, we show that a confusion between coordinate 
lengths and proper lengths in fact disqualifies the model in Ref.~\cite{KIII} from being traversable. 
It turns out that for this model, radially infalling particles and light rays 
reach the throat only after an infinite lapse of affine parameter, due to 
the extremely slow flaring of the wormhole throat. Related constraints were 
also derived for two of Kuhfittig's earlier models. One lesson to be drawn from our results 
is that simply concentrating the exotic matter, in a classical 
analysis, to an arbitrarily small region around the wormhole throat is, by itself, 
not sufficient to guarantee both traversability and consistency with (or evasion of) the 
quantum inequality bounds.  

\vskip 0.2in   

\centerline{\bf Acknowledgments}  
The authors would like to thank Matt Visser and Larry Ford for helpful
discussions, and Lutz Osterbrink for comments on the manuscript. 
This work was supported in part by the National
Science Foundation under Grant PHY-0139969 (to TAR). One of us (TAR) would like 
to thank the Mathematical Physics Group at the University of York for their kind 
hospitality during the course of some of this work.


\begin{thebibliography}{--}
\bibitem{QEIs} Sometimes QIs are also called ``quantum energy
inequalities'' (QEIs); more specific terms such as ``quantum weak energy
inequalities'' are also used.
\bibitem{widthnote} Any measure of the width would do, e.g., the
standard deviation, or the length of the smallest interval outside which
$q$ vanishes.
\bibitem{Accn_comment} The same would apply to general timelike
trajectories provided the sampling timescale is also
small in relation to the acceleration timescale $|a^b a_b|^{-1}$ [of
course the latter timescale is infinite for a geodesic observer].
\bibitem{MVTnote} In Sec. 11.4 of Ref.~\cite{VBOOK}, it is claimed that
this argument proves that
NEC is violated throughout some interval $(r_0,r_*)$ near the throat.
This need not be true: a $C^1$ function $f(x)$ can vanish at $x=0$ and
be strictly positive in $x>0$ without having $f'(x)>0$ on some interval
$(0,x_*)$. An example is provided by $f(x)=x^3+x^4+(x^3-x^4)\sin(1/x)$.
\bibitem{VBOOK-ZD} These are essentially the same as the zero density 
wormholes discussed in Sec. 13.4.2 of Ref.~\cite{VBOOK}.
\bibitem{fnote} Note that refs.~\cite{Fewster} and~\cite{FewR} use
different conventions for the Fourier transform.
\end{thebibliography}
\end{document}